\documentclass[iop]{emulateapj}
\usepackage{apjfonts}
\newcommand{\be}{\begin{equation}}
\newcommand{\ee}{\end{equation}}
\newcommand{\bea}{\begin{eqnarray}}
\newcommand{\eea}{\end{eqnarray}}
\newcommand{\Msun}{M_{\odot}}

\def\npix{N_{\rm pix}}

\def\gband{g_{475}}
\def\iband{I_{814}}

\shortauthors{CONROY \& VAN DOKKUM}
\shorttitle{Pixel Color Magnitude Diagrams}

\begin{document}

%---------------------------------------------------------%
\title{Pixel Color Magnitude Diagrams for Semi-Resolved Stellar
  Populations: The Star Formation History of Regions within the Disk
  and Bulge of M31}
%---------------------------------------------------------%

\author{Charlie Conroy\altaffilmark{1} \&
  Pieter G. van Dokkum\altaffilmark{2}}

\altaffiltext{1}{Department of Astronomy, Harvard University,
  Cambridge, MA, USA}
\altaffiltext{2}{Department of Astrophysical
  Sciences, Yale University, New Haven, CT, USA}

\slugcomment{Submitted to ApJ}

\begin{abstract}

  The analysis of stellar populations has, by and large, been
  developed for two limiting cases: spatially-resolved stellar
  populations in the color-magnitude diagram, and integrated light
  observations of distant systems.  In between these two extremes lies
  the semi-resolved regime, which encompasses a rich and relatively
  unexplored realm of observational phenomena.  Here we develop the
  concept of pixel color magnitude diagrams (pCMDs) as a powerful
  technique for analyzing stellar populations in the semi-resolved
  regime.  pCMDs show the distribution of imaging data in the plane of
  pixel luminosity vs. pixel color.  A key feature of pCMDs is that
  they are sensitive to all stars, including both the evolved giants
  and the unevolved main sequence stars.  An important variable in
  this regime is the mean number of stars per pixel, $\npix$.
  Simulated pCMDs demonstrate a strong sensitivity to the star
  formation history (SFH) and allow one to break degeneracies between
  age, metallicity and dust based on two filter data for values of
  $\npix$ up to at least $10^4$.  We extract pCMDs from {\it Hubble
    Space Telescope (HST)} optical imaging of M31 and derive
  non-parametric SFHs from $10^6$ yr to $10^{10}$ yr for both the
  crowded disk and bulge regions (where $\npix\approx30-10^3$).  From
  analyzing a small region of the disk we find a non-parametric SFH
  that is smooth and consistent with an exponential decay timescale of
  4 Gyr.  The bulge SFH is also smooth and consistent with a 2 Gyr
  decay timescale.  pCMDs will likely play an important role in
  maximizing the science returns from next generation ground and
  space-based facilities.

\end{abstract}


%---------------------------------------------------------%

\section{Introduction}
\label{s:intro}

Analysis of the stellar populations of both star clusters and galaxies
has provided the foundation for much of our understanding of the
formation and evolution of galaxies.  Interpreting observations of
complex stellar populations relies on two key ingredients - stellar
evolution models and stellar spectral libraries.  The task of
comparing observations to models reduces fundamentally to the task of
searching for linear combinations of these two key ingredients
(usually cast as the star formation history, SFH, and metallicity, Z),
often with the inclusion of additional ingredients such as dust
attenuation and emission.

Despite these underlying common ingredients, the analysis of stellar
populations in practice is almost universally treated in one of two
limiting cases.  In the first case, fully resolved stellar populations
are available, where one is able to robustly measure fluxes of each
star in the system (down to some limit).  One then models these
observations in the color magnitude diagram (CMD) by fitting stellar
evolution models to the data \cite[e.g.,][]{Dolphin02, Tolstoy09}.  In
the second case, fully unresolved stellar populations are assumed,
e.g., when observing distant galaxies.  In this case all of the stars
add together to contribute to an integrated spectrum which is then
modeled with stellar population synthesis techniques
\citep[e.g.,][]{Walcher11, Conroy13b}.  Poisson fluctuations are
generally negligible and one can assume a fully populated initial mass
function (IMF). 

A useful way to describe these two extremes within a common framework
is by considering the mean number of stars per pixel,
$\npix$.\footnote{For simplicity we characterize the pCMDs in terms of
  $\npix$ but in reality the key variable is the number of stars per
  resolution element.}  In the case of fully resolved stellar
populations this number is very small, e.g., $\lesssim10^{-2}$.  In
contrast, distant galaxies are typically in the regime where
$10^6\lesssim\npix\lesssim10^{12}$.  The intermediate, `semi-resolved'
regime where $1\lesssim\npix\lesssim10^6$, is relatively unexplored
territory.  Examples at the upper end of this regime includes surface
brightness fluctuations \citep[SBF;][]{Tonry88}, fluctuation
spectroscopy \citep{vanDokkum14b}, pixel-level time variability due to
the finite number of long period variables per pixel
\citep{Conroy15a}, and `dis-integrated' light analysis of stellar
halos \citep{Mould12}.  At the lower end of this regime,
\citet{Beerman12} demonstrated that strong constraints on the ages of
low mass star clusters could be obtained by analyzing the integrated
light of the unevolved main sequence stars after isolating and
removing the rare luminous evolved stars.

Even with the exquisite angular resolution and PSF stability delivered
by {\it Hubble Space Telescope (HST)} imaging, the main body of every
galaxy beyond the nearest dwarfs and out to at least the Virgo cluster
are in the semi-resolved regime.  For example, {\it HST} imaging of
M31, obtained by the Panchromatic Hubble Andromeda Treasury (PHAT)
Survey is crowding-limited well above the oldest main sequence turnoff
point across the entire disk of M31 \citep{Dalcanton12}.  Producing a
complete photometric catalog down to the oldest main sequence turnoff
point is considered the gold standard in resolved star analysis
because the turnoff point is the most reliable age indicator, and
hence resolving the oldest main sequence turnoff point enables the
most precise and accurate SFHs for the entire age range.  In spite of
this crowding, strong constraints on the detailed SFH can still be
obtained in such regions by modeling the upper main sequence and
evolved giants that are above the crowding limit
\citep[e.g.,][]{Lewis15, Williams15}.  The situation in the bulge of
M31 is much worse - the data are so crowded in the optical and NIR
that point source photometry of even the brightest giants is difficult
to interpret without extensive simulations to map the completeness and
bias resulting from crowding \citep{Dalcanton12, Williams14}. This is
not surprising - the mean number of stars per pixel in the bulge is
$\sim10^2-10^3$ at {\it HST} resolution.

The goal of this paper is to develop a new analysis framework that
enables seamless transition from the fully resolved to fully
unresolved regimes.  The basic idea is to forego point source
photometry and instead construct pixel CMDs (pCMDs) directly from the
imaging data.  This approach requires generating models in the image
plane, which entails creating complex stellar populations
pixel-by-pixel and then convolving the model image with the PSF.  In
many other respects the analysis procedure is similar to modeling
resolved stellar populations, in which one converts the photometric
catalog into a Hess diagram in CMD space.

We expect that pCMDs will be an important tool for analyzing future
data.  On the near horizon are an array of new facilities,
instruments, and observatories, including the {\it James Webb Space
  Telescope (JWST)}, WFIRST, Euclid, the Large Synoptic Survey
Telescope (LSST), and three 30m class ground-based telescopes (ELTs).
While these facilities will revolutionize many aspects of
extragalactic astrophysics, it is important to underline the fact that
almost none of these facilities will deliver significantly better
spatial resolution than {\it HST}.  The only possibility for real gain
is with the ELTs, provided that they can deliver diffraction-limited
imaging over wide fields with a highly stable PSF \citep[see
e.g.,][for the expected gains in the limit of a perfectly known and
stable PSF]{Olsen03, Greggio12, Schreiber14}.  Even in this case the
regime of semi-resolved populations will simply be pushed to somewhat
larger distances, such as the Virgo and Coma clusters.  In order to
maximize returns from these new facilities, we must therefore develop
new tools specifically for the semi-resolved universe.

This paper is organized as follows.  In Section \ref{s:crowd} we
provide a brief review of the crowding limit.  Section \ref{s:models}
describes the modeling of pCMDs and in Section \ref{s:fit} we describe
our approach to fitting models to data in pCMD space.  Section
\ref{s:data} contains a comparison between models and observations of
M31.  A discussion and summary are provided in Sections \ref{s:disc}
and \ref{s:sum}.  Where necessary we adopt the AB magnitude zeropoint
system \citep{Oke83}.

%---------------------------------------------------------%

\section{The Crowding Limit}
\label{s:crowd}

Crowding of sources will limit the flux at which one can reliably
estimate photometry.  This crowding, or confusion limit, has been
studied extensively in both the radio and optical astronomy
communities \citep[e.g.,][]{Scheuer57, Condon74, Renzini98, Hogg01,
  Olsen03}.  \citet{Olsen03} provided detailed simulations of the
crowding limit for realistic stellar populations as a function of many
of the controlling parameters (age, metallicity, IMF, wavelength).
Careful analysis of real data requires measuring the crowding limit
across the field by injecting artificial stars into the image
\citep[e.g.,][]{Dalcanton12, Williams14}.  In this paper we are only
interested in an approximate crowding limit in order to provide
perspective on the delineation between the resolved and semi-resolved
regime.  For this purpose we adopt the rule of thumb which states that
photometry becomes confusion-limited when the background flux level is
equal to that which would be produced if light from the source were
spread out over 30 resolution elements \citep{Hogg01}.  In our case we
assume that one resolution element is equal to 10 {\it HST} ACS
pixels, as 10 pixels contain $\approx60$\% of the total flux.  We have
compared this simple estimate of the crowding limit to the models
presented in \citet{Olsen03} and find that the simple rule of thumb
reproduces the predictions from the detailed models generally to
within 0.5 mag, which, for the purposes of guiding the eye, is
sufficient.

\begin{figure}[!t]
\center
\includegraphics[width=0.47\textwidth]{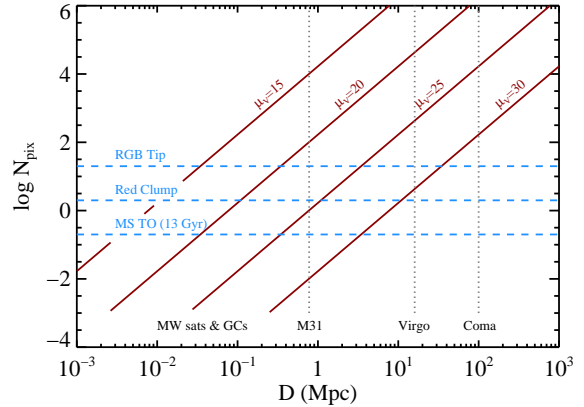}
\vspace{0.1cm}
\caption{Relation between the mean number of stars per pixel, $\npix$,
  distance, and surface brightness for a pixel scale of 0.05\arcsec
  and a resolution element of 10 pixels.  Also shown are the
  approximate crowding limits for three key stellar phases: the oldest
  main sequence turnoff (MS TO), the red clump, and the tip of the
  RGB.  Approximate locations of a variety of objects are included at
  the bottom of the figure.  The intersection between the dashed and
  solid lines marks the region above which that particular stellar
  phase is crowding limited.  For example, at the distance of M31,
  regions with a surface brightness of 25 mag arcsec$^{-2}$ fully
  resolve the RGB tip, barely resolve the red clump, and do not
  resolve the MS TO.  }
\label{fig:npix}
\end{figure}

The $\npix$ parameter depends on the distance, surface brightness
(which is proportional to the stellar surface density), mass-to-light
ratio, and the IMF.  Figure \ref{fig:npix} shows the relation between
$\npix$, distance, and surface brightness.  We have assumed a
mass-to-light ratio of 4.0 and a mass-to-number ratio of 2.0 in order
to convert luminosities into numbers, and a pixel resolution element
of $0.05\arcsec$ (i.e., the pixel scale of ACS).  Also shown in this
figure is the approximate crowding limit for three important stellar
phases: the main sequence turn-off at 13 Gyr, the red clump, and the
tip of the red giant branch (RGB).  Above these limits in $\npix$ the
phases are crowding limited, which means that stars in this phase, and
fainter, cannot be reliably separated from the background.  Distances
to a variety of stellar systems including Milky Way satellite galaxies
and globular clusters (GCs), M31, and the Virgo and Coma clusters are
included at the bottom of the figure.  As mentioned in the
Introduction, most of the next-generation facilities, including WFIRST
and JWST, will not deliver significant improvements in spatial
resolution compared to {\it HST}, and so we can expect the landscape
depicted in Figure \ref{fig:npix} to remain largely unchanged for the
next $10-20$ years.

%---------------------------------------------------------%

\section{Pixel Color Magnitude Diagrams}
\label{s:models}

\begin{figure*}[!t]
\center
\includegraphics[width=0.9\textwidth]{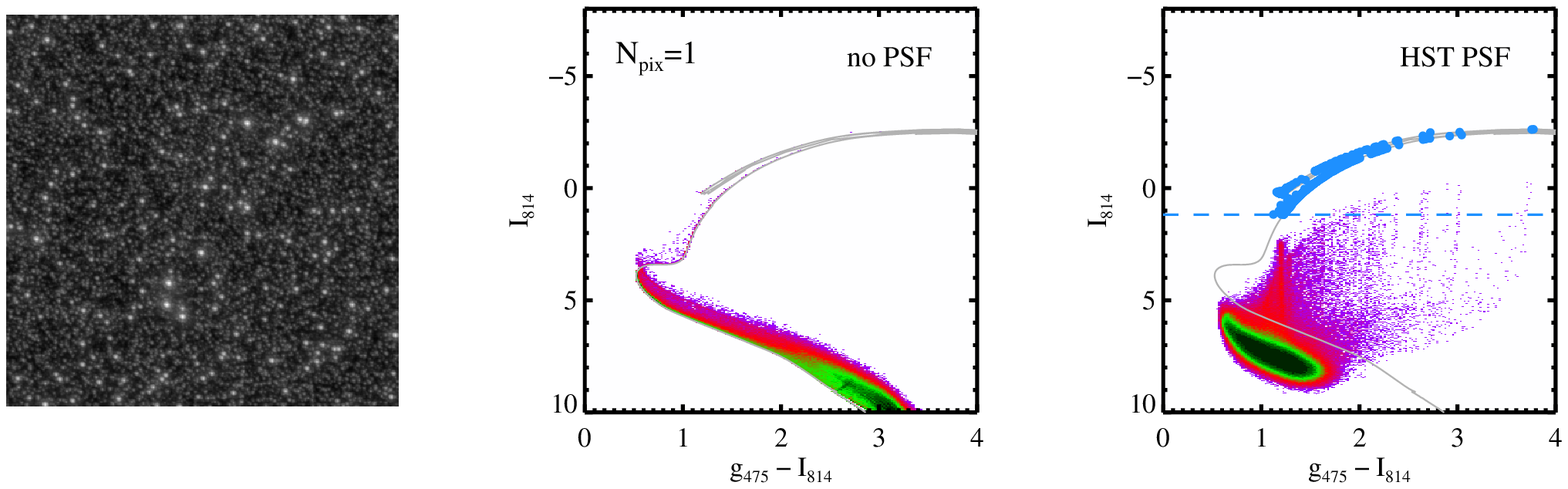}
\includegraphics[width=0.9\textwidth]{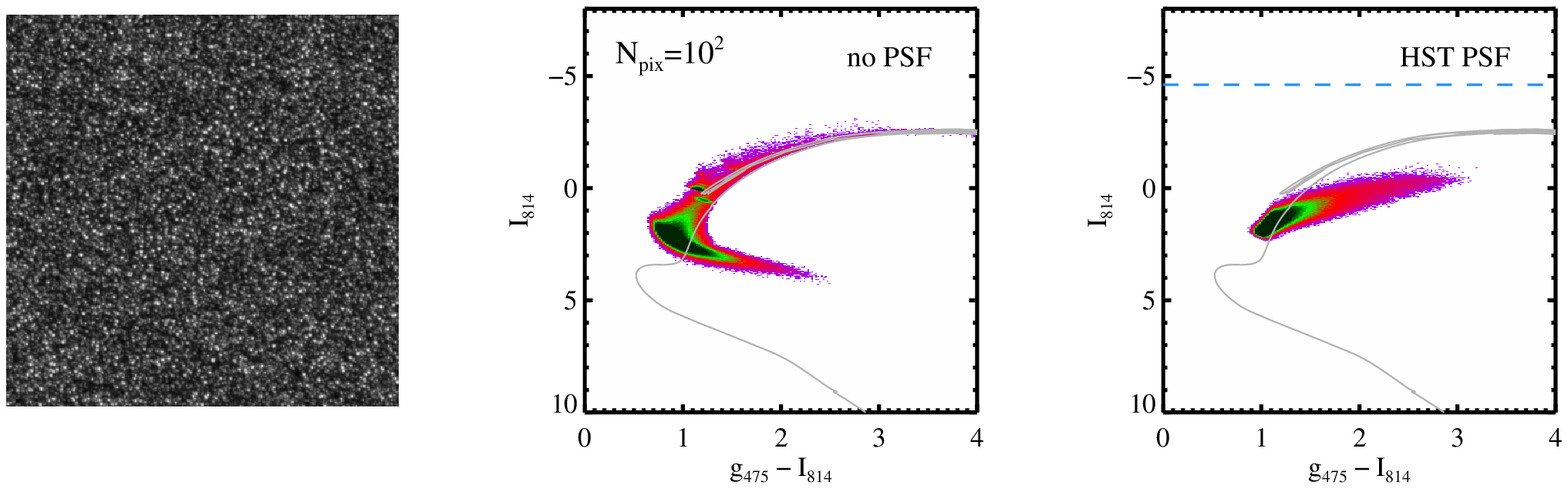}
\includegraphics[width=0.9\textwidth]{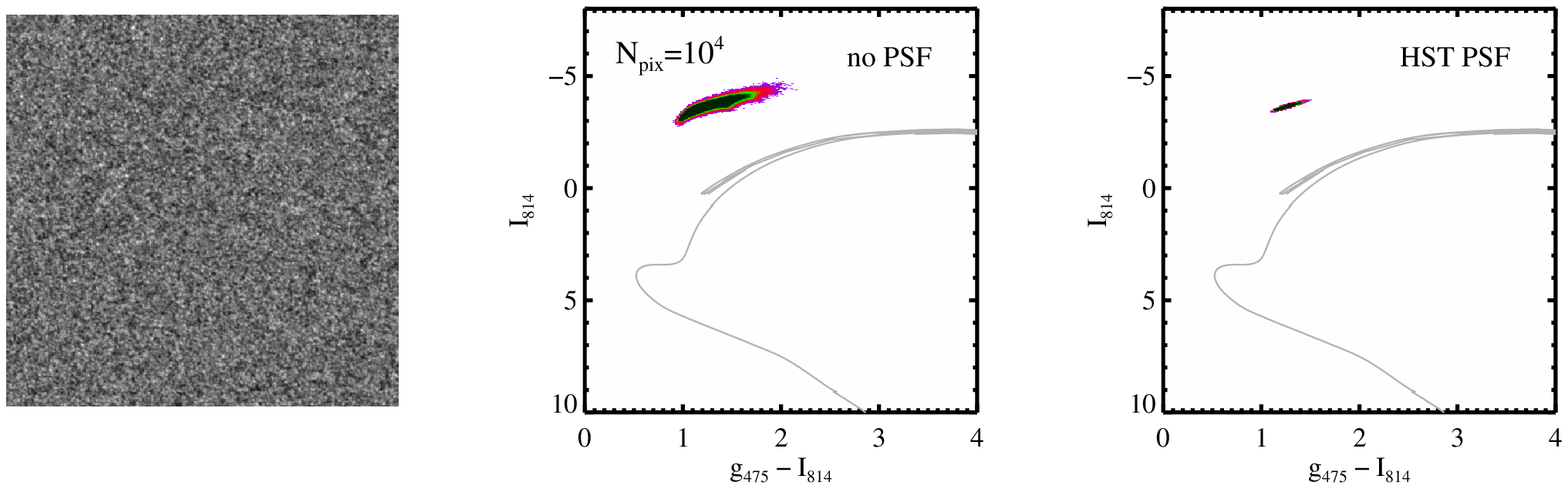}
\vspace{0.1cm}
\caption{Model images and pCMDs as a function of the number of stars
  per pixel, $\npix$.  All models adopt a solar metallicity, 10 Gyr
  stellar population (shown as a grey line in each pCMD to guide the
  eye).  Simulated pCMDs contain $1024^2$ pixels; $400^2$ pixel
  subregions are shown in the PSF-convolved images in the left panels.
  Middle and right panels show the resulting pCMDs without and with
  convolution with the {\it HST} ACS F814W PSF.  The pCMDS are
  displayed as Hess diagrams with a logarithmic color mapping.
  Horizontal dashed lines in the right panels mark the approximate
  crowding limit, above which PSF photometry is possible.  In the
  upper right panel, blue symbols denote what would be recoverable
  from PSF photometry; there are 1062 stars above the crowding limit
  for $\npix=1$ but none for $\npix\ge10^2$.}
\label{fig:overview}
\end{figure*}

\subsection{Methods}

pCMDs are constructed by simply measuring magnitudes within a pixel
and plotting those pixel magnitudes in a color magnitude diagram.
Obviously this requires the pixels from different images, bands,
etc. to be registered to the exact same reference frame and to have
the same pixel scale.  The modeling of stellar populations in pCMD
space is in principle very straightforward.  Our goal is to model the
image plane, and from that image construct pCMDs.  In order to model
the image plane we must create stellar populations {\it at each pixel}
and then convolve the model image with the PSF.  At each pixel we draw
stars according to weights specified by the product of the IMF, the
star formation history (SFH), and the mean number of stars per pixel.
The flux of each star can be reddened according to a reddening law.
The stars within a pixel are summed together to produce the final
spectral energy distribution of the pixel.

In practice we adopt a Salpeter IMF \citep{Salpeter55} with a
lower-mass cutoff of $0.08\Msun$ and isochrones from the MIST
project \citep{Choi16}.  The SFH is set by weights supplied in 7 age
bins and can either be specified non-parametrically or via parametric
relations (a constant model or an exponential model with a decay
timescale denoted by $\tau_{\rm SF}$).  Note that the SFH extends from
$10^6$ yr to $10^{10}$ yr.  We allow for a single reddening parameter,
$E(B-V)$, that is applied to all stars equally, with an $R_V=3.1$
reddening law from \citet{Schlafly11}.  Bandpass filters are adopted
from the {\it HST} ACS camera and for brevity we refer to the {\it
  HST} filters with the following abbreviations: $\gband=$F475W and
$\iband=$F814W.  We adopt a single PSF derived from ACS F814W imaging
when convolving images (the F475W and F814W PSFs are very similar and
so this simplification is unlikely to impact our comparison to
observations in later sections).

In most respects the modeling process is similar to modeling resolved
stellar populations.  However, there are key differences: first, since
we are modeling the image plane, we must take into account the PSF in
the modeling procedure.  Second, computational considerations require
us to model a finite image plane, which implies that the model
contains a non-trivial stochastic noise component (we return to this
point in Section \ref{s:fit}).  Third, the key variable $\npix$ must
be modeled.  In reality one might need to consider a distribution
function for $\npix$ rather than a single value.  A simplifying
feature of modeling in pCMD space is that one need not explicitly
consider stellar binarity since we are not attempting to resolve
individual stars.  For these reasons the number of free parameters is
not necessarily much larger, although the modeling of populations in
pCMD space is substantially more computationally intensive.

\begin{figure}[!t]
\center
\includegraphics[width=0.45\textwidth]{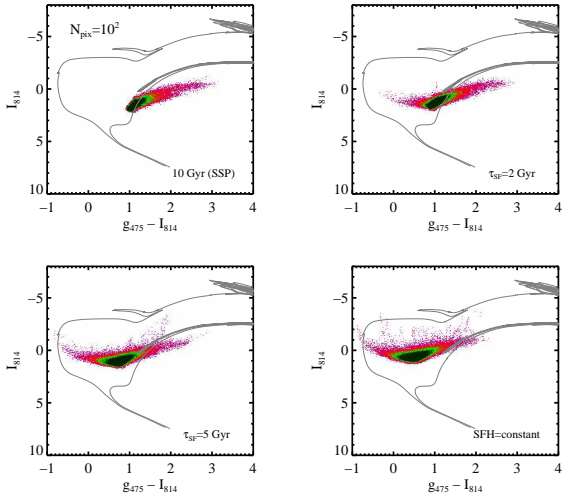}
\vspace{0.1cm}
\caption{Sensitivity of pCMDs to the star formation history.  Solar
  metallicity models with $\npix=10^2$ are shown for four SFHs: a 10
  Gyr single-age model, $\tau$-model SFHs with $\tau_{\rm SF}=2$ and 5
  Gyr, and a constant SFH.  The pCMDS are displayed as Hess diagrams
  with a logarithmic color mapping.  Isochrones at 0.01 and 10 Gyr are
  shown in grey to guide the eye. There is clear sensitivity to the
  different age components in these SFHs in the mean color of the
  faint pixels, the relative numbers of RGB and upper main sequence
  stars, and the distribution of stars along the upper main sequence.}
\label{fig:varysfh1}
\end{figure}

\subsection{Information Content of pCMDs}

In this section we illustrate the information content of pCMDs as a
function of $\npix$, SFH, and metallicity.

We begin with Figure \ref{fig:overview}, which shows the image plane
and pCMDs as a function of $\npix$ in the $\gband$ and $\iband$
filters.  We show pCMDs both with and without convolution with the PSF
in order to illustrate the critical role played by the PSF.  Models
were generated for solar metallicity 10 Gyr single-age stellar
populations with zero reddening.  The distribution of points in the
pCMD is represented by a Hess diagram with a logarithmic color
mapping. Some of the discrete features evident in the top right panel
are the result of the finitely-sampled isochrone, PSF and image-plane
(by $\npix\gtrsim10^2$ these numerical issues become negligibly
small).  

One sees clearly from this figure that the morphology in the pCMD
varies smoothly from $\npix=1$ to $\npix=10^4$, indicating that the
information content in these diagrams also varies smoothly.  The
approximate crowding limit is shown in the PSF-convolved pCMDs in
order to underline the rich morphology that lies in the regime that is
not included in resolved stellar population analysis.  We also show in
the top right panel the stars that are recoverable as resolved sources
above the crowding limit.  By $\npix\ge10^2$ there are no stars in the
entire image above the (approximate) crowding limit.

\begin{figure}[!t]
\center
\includegraphics[width=0.45\textwidth]{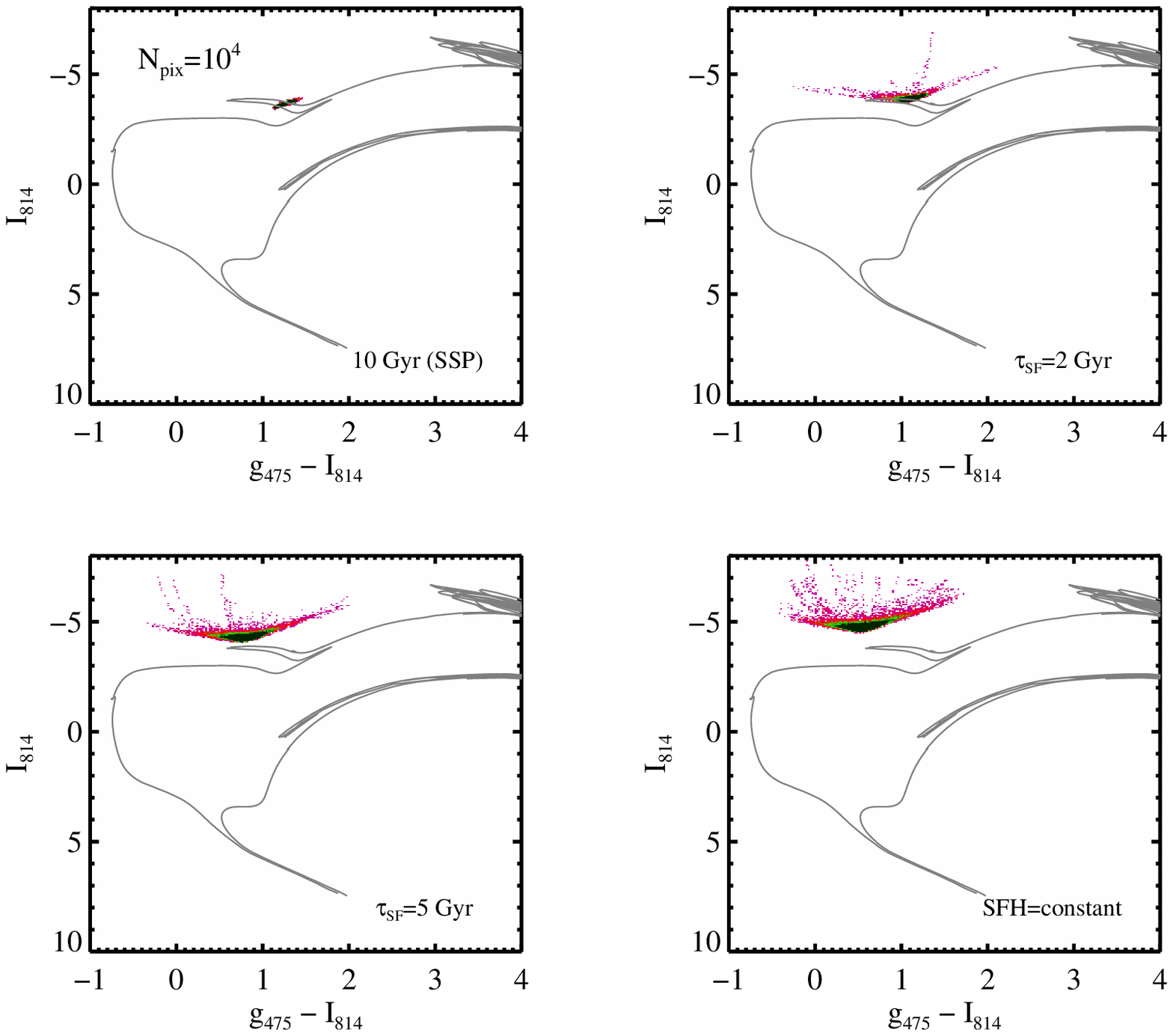}
\vspace{0.1cm}
\caption{Same as Figure \ref{fig:varysfh1}, now for $\npix=10^4$.}
\label{fig:varysfh3}
\end{figure}

Figures \ref{fig:varysfh1} and \ref{fig:varysfh3} show the effect of
different SFHs in pCMD space for two values of $\npix$.  In each
figure we compare a single age population at 10 Gyr to two
exponentially-declining ($\tau$-model) SFHs with timescales of
$\tau_{\rm SF}=2$ and 5 Gyr, and a constant SFH.  All models are
reddening-free and at solar metallicity.  Several important trends are
evident.  First, there is clearly a varying balance between luminous
red and blue pixels, due to the varying influence of upper main
sequence and RGB stars.  Second, the distribution of stars along the
upper main sequence is clearly changing with the SFH.  Third, the mean
color of the faintest pixels and the faint limit of the faintest
pixels also varies with SFH.  Also notice that some of the finger-like
features extending vertically toward brighter fluxes are the result of
one or a few rare bright stars.  In pixel space a single bright star
will occupy many pixels owing to the effect of the PSF.  These pixels
will vary in brightness at approximately constant color (depending in
detail on the level of similarity of the PSFs of the two filters).  In
many cases those finger-like structures would be recoverable as
resolved sources above the crowding limit.

\begin{figure}[!t]
\center
\includegraphics[width=0.45\textwidth]{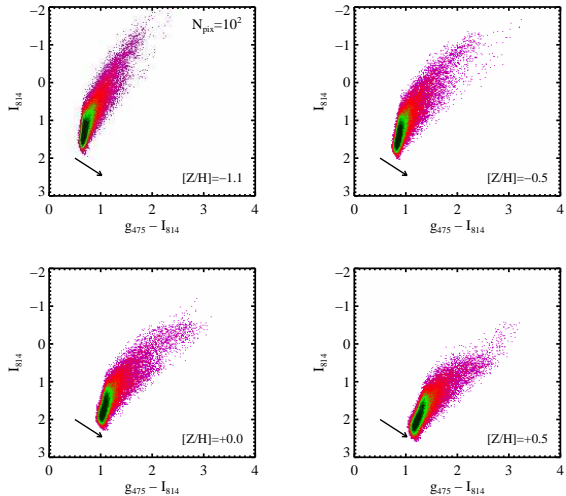}
\vspace{0.1cm}
\caption{pCMDs as a function of metallicity.  Each panel shows a solar
  metallicity 10 Gyr single-age model displayed as a Hess diagram
  (color mapping is logarithmic).  Also shown is a reddening vector
  for a standard $R_V=3.1$ reddening law and $E(B-V)=0.3$.  While both
  metallicity and reddening result in overall redder colors, the
  morphology of the pCMD contains a significant amount of metallicity
  information. For example the pCMDs transition from being more
  vertically-aligned to more horizontally-aligned with increasing
  metallicity.  It should therefore be possible to separately
  constrain metallicity and reddening with only two band imaging. }
\label{fig:mpix_z}
\end{figure}

Figure \ref{fig:mpix_z} shows pCMDs as a function of metallicity.
Models were generated for a 10 Gyr single-age solar metallicity
population with zero reddening.  Also shown is a reddening vector
assuming the reddening law of \citet{Schlafly11} for a reddening of
$E(B-V)=0.3$ and $R_V=3.1$.  As is well-known, an increase in dust or
metallicity will result in redder colors.  In unresolved data it is
generally impossible to separate the effects of dust, metallicity, and
age with a single color.  However, with semi-resolved data there is
clearly much more information.  Notice that while the pCMDs become
redder overall with increasing metallicity, the morphology of the data
in pCMD space also changes, becoming more horizontally-aligned with
increasing metallicity.  We therefore expect that pCMDs will enable
the disentangling of metallicity and dust effects even when only two
imaging bands are available.  Though not shown, we have made similar
diagrams for $\iband-H_{160}$ colors and find an even stronger
sensitivity to metallicity, suggesting that optical-NIR colors may be
better suited for jointly estimating metallicities and reddening.

%---------------------------------------------------------%

\section{Fitting Data in PCMD Space}
\label{s:fit}

In this section we describe our approach to fitting data in pCMD space
and demonstrate its effectiveness by testing against mock observations.

\begin{figure*}[!t]
\center
\includegraphics[width=0.8\textwidth]{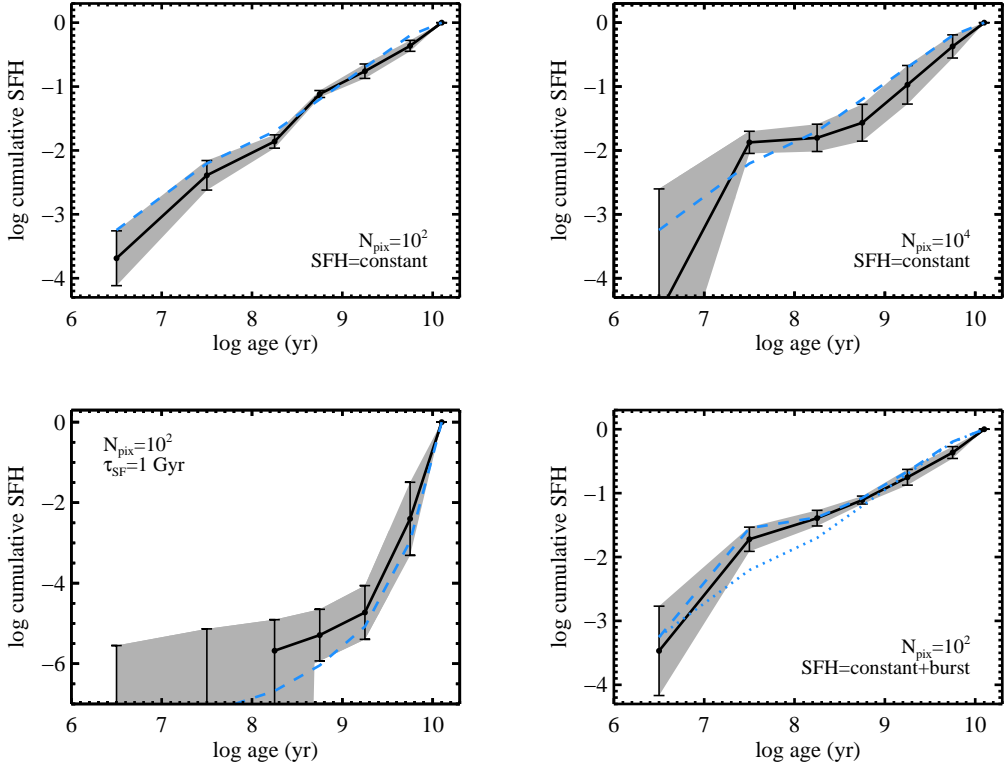}
\vspace{0.1cm}
\caption{SFH recovery tests with mock data.  In each panel the
  recovered SFH (solid line) is compared to the input SFH (dashed
  line).  Upper panels show the recovered SFH and $1\sigma$ errors for
  mock data with a constant SFH and solar metallicity.  The upper left
  panel shows results for $\npix=10^2$ and the upper right panel shows
  results for $\npix=10^4$.  In both cases the total mass of the
  population is fixed, so the total number of pixels in the upper
  right panel is $100\times$ fewer than in the left.  The bottom left
  panel shows the recovered SFH for an input model with a $\tau-$model
  SFH with $\tau_{\rm SF}=1$ Gyr.  The bottom right panel shows
  results for a constant SFH with a burst in the second youngest age
  bin resulting in $5\times$ more mass.  Also shown in this panel is a
  constant SFH (dotted line) for comparison. }
\label{fig:fitmock}
\end{figure*}

\subsection{Fitting Technique}

The basic framework for fitting observations in pCMD space is as
follows.  One must simulate an image plane (actually two, one for each
band) of some dimension for a given input SFH, reddening, and $\npix$,
convolve with the appropriate PSF, add observational errors, bin into
a Hess diagram in pCMD space, compute the likelihood of this model
given the data, and iterate.  We implement this in practice in the
following manner.  The image plane is simulated at a resolution of
$256^2$ pixels.  The SFH has a free, non-parametric form discretized
into 7 age bins with the following boundaries:
$(6.0,7.0,8.0,8.5,9.0,9.5,10.0,10.2)$ in units of log age (yr).  We
assume that the SFR is constant within each bin.  The mass in each
bin, $M_i$, constitutes 7 free parameters ($\npix$ is derived from the
integral of the SFH).  The reddening is taken to be a single
parameter: log $E(B-V)$.  The final free parameter is the metallicity,
[Z/H].  We fit for a single metallicity by interpolating within the
isochrone tables (which include the bolometric corrections).  There
are thus 9 free parameters in total.  The priors on the parameters are
flat in log space over the boundaries: [Z/H]$=(-1.1,0.5)$, log
$E(B-V)=(-6.0,0.0)$, log $M_i/M_{\rm tot}=(-10.0,0.0)$.  The lower
limit on the metallicity was simply a practical consideration given
that we are interested in this paper in metal-rich regions within M31.

The PSF is implemented by dividing the image into 16 subregions.
Within each subregion the image is convolved with a PSF that has been
shifted by 1/4 of a pixel.  We do this because we have assumed in the
model that the stellar populations reside at the center of each pixel,
whereas in reality the stars can of course fall anywhere in the image
plane, not only at pixel centers.

Observational (photon counting) errors are applied to the model by
converting the pixel fluxes into counts by specifying a distance to
the object of interest and an exposure time.  At each pixel we then
draw from a Poisson distribution with a mean given by the mean number
of counts in each pixel.

The default MIST isochrones are very densely sampled in mass at each
age \citep[see][for details]{Dotter16}.  In order to ease the
computational burden in the fitting we have made use of isochrones
that are sampled with $5\times$ fewer equivalent evolutionary points,
resulting in $\sim100-200$ points per isochrone.  In testing we have
found that this reduction in isochrone points has a very minor effect
on the resulting pCMD, mostly affecting the very rare and luminous
stars, which in any event do not receive significant weight in the
fit.

Errors on both the data and the model Hess diagrams are computed
directly from the number of pixels at each point (assuming Poisson
statistics).  Both the model and data Hess diagrams are normalized to
unity.

We begin the fitting procedure by fitting only for $\npix$ and a
smooth exponentially-declining SFH specified by $\tau_{\rm SF}$.  We
fix the reddening to zero and the metallicity to solar.  This two
parameter fit provides a good starting position for the main fitting
routine, which employs the Markov chain Monte Carlo (MCMC) technique
\texttt{emcee} \citep{Foreman-Mackey13}.  We use 256 walkers and find
that the solution is typically well-converged after $10^3$ iterations.

Fitting pCMD data presents several unique challenges not encountered
when fitting resolved CMD data.  Chief among them is the Poisson
nature of drawing stars and populating them in a finite image.  For
$256^2$ image pixels and $N_{\rm iso}\approx 3000$ isochrone points
this requires $\sim10^8$ Poisson draws per likelihood call.  This is
computationally very expensive.  Moreover, the stochastic nature of
the model implies that the exact same set of parameters will result in
a somewhat different model, and hence a different $\chi^2$ value.
Such a model would require more sophisticated search techniques than
standard MCMC.  This latter issue could be mitigated by simulating an
image of sufficiently large number of pixels, but in our testing even
$1024^2$ pixels was not sufficient to overcome the Poisson noise.  In
order to circumvent these two issues (the expense of drawing many
Poisson numbers and the stochastic nature of the model), we decided to
make the following approximation.  Rather than drawing random numbers
for the computation of each Poisson draw, we created a fixed set of
$256^2\times N_{\rm iso}$ random numbers at the beginning of the
program.  This ensures that each isochrone point has a unique random
number at each image pixel, and it also guarantees that the model is
deterministic and is computationally faster by about a factor of 3.
We then fit each dataset 10 times with a different random number seed
and combine the 10 posteriors in an attempt to account for the
uncertainties in the model induced by stochastic effects (note that
when we generate mock data, we use a random number seed different the
10 used in the fitting).

We make two additional simplifications.  We assume that isochrone
points with a mass $<0.7\Msun$ or a mean number (which is the product
of the IMF and SFH weights) $\langle N\rangle>10^3$ contain no
pixel-to-pixel variation and hence are not discretely drawn.  For the
rest, we draw from a Poisson distribution if $\langle N\rangle<100$
and a Gaussian distribution otherwise.  With these simplifications
each likelihood call takes $\approx1$ s.

As this is the first attempt to fit observations in pCMD space, we
have taken a somewhat simplified approach to fitting the data.  Areas
for future improvement include the following: 1) exploring techniques
for rapidly generating models that do not suffer from stochastic
effects, e.g., by simulating images of much larger numbers of pixels;
2) allowing for more metallicity components, either in the form of a
metallicity distribution function at a fixed age or the freedom for
each age component to have its own metallicity; 3) a dust model
characterized by more than a single $E(B-V)$ value, e.g., allowing for
a distribution function of reddening values as in \citet{Dalcanton15};
4) fitting not just $\npix$ but $P(\npix)$, i.e., allowing for the
fact that a given physical region will in nearly all cases have a
distribution of $\npix$ values; 5) a more detailed modeling of the
PSF, including e.g., its spatial variation across the image.  All of
these improvements are straightforward to implement although most will
significantly increase the computational expense of each likelihood
evaluation.

\begin{figure*}[!t]
\center
\plottwo{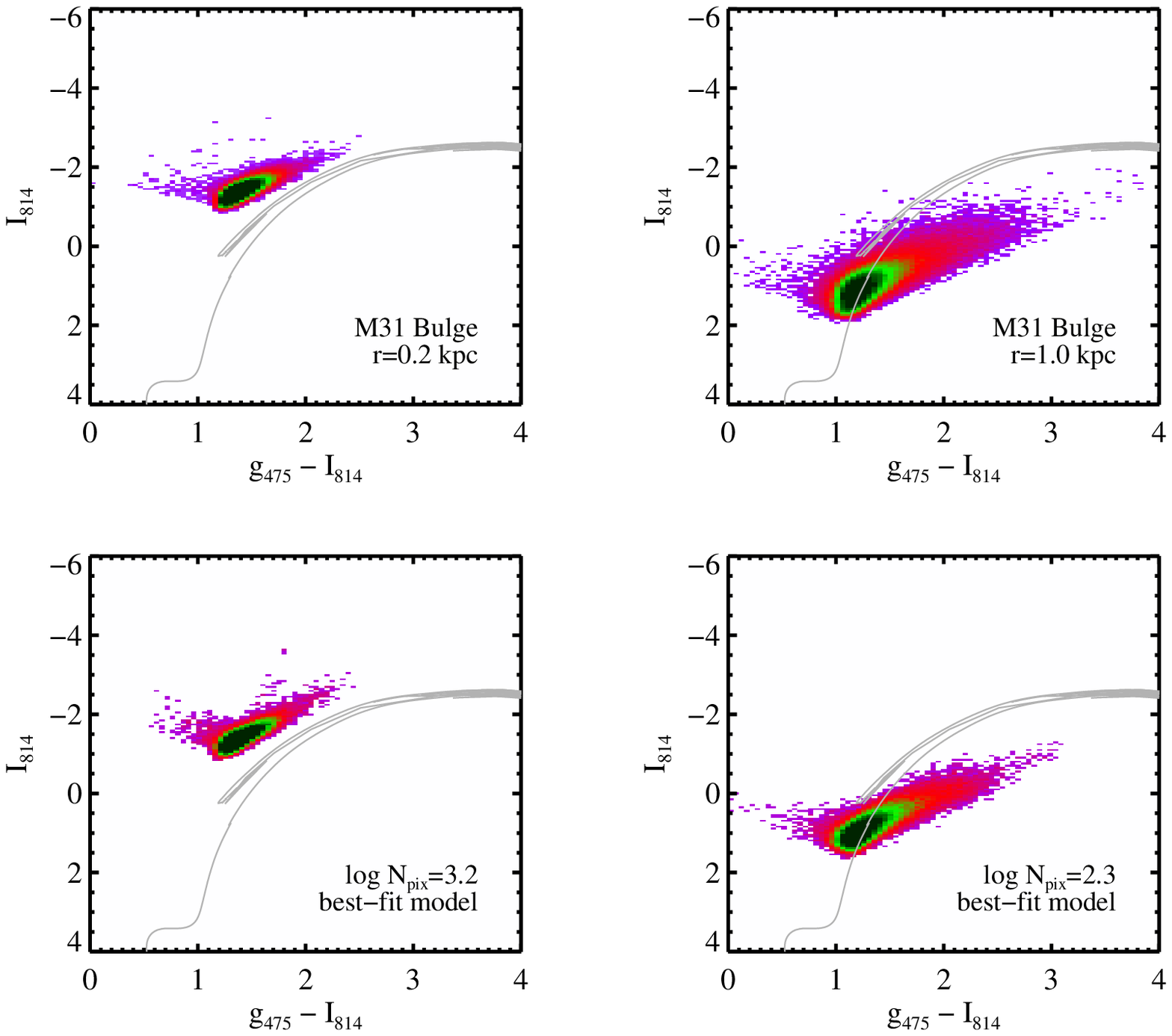}{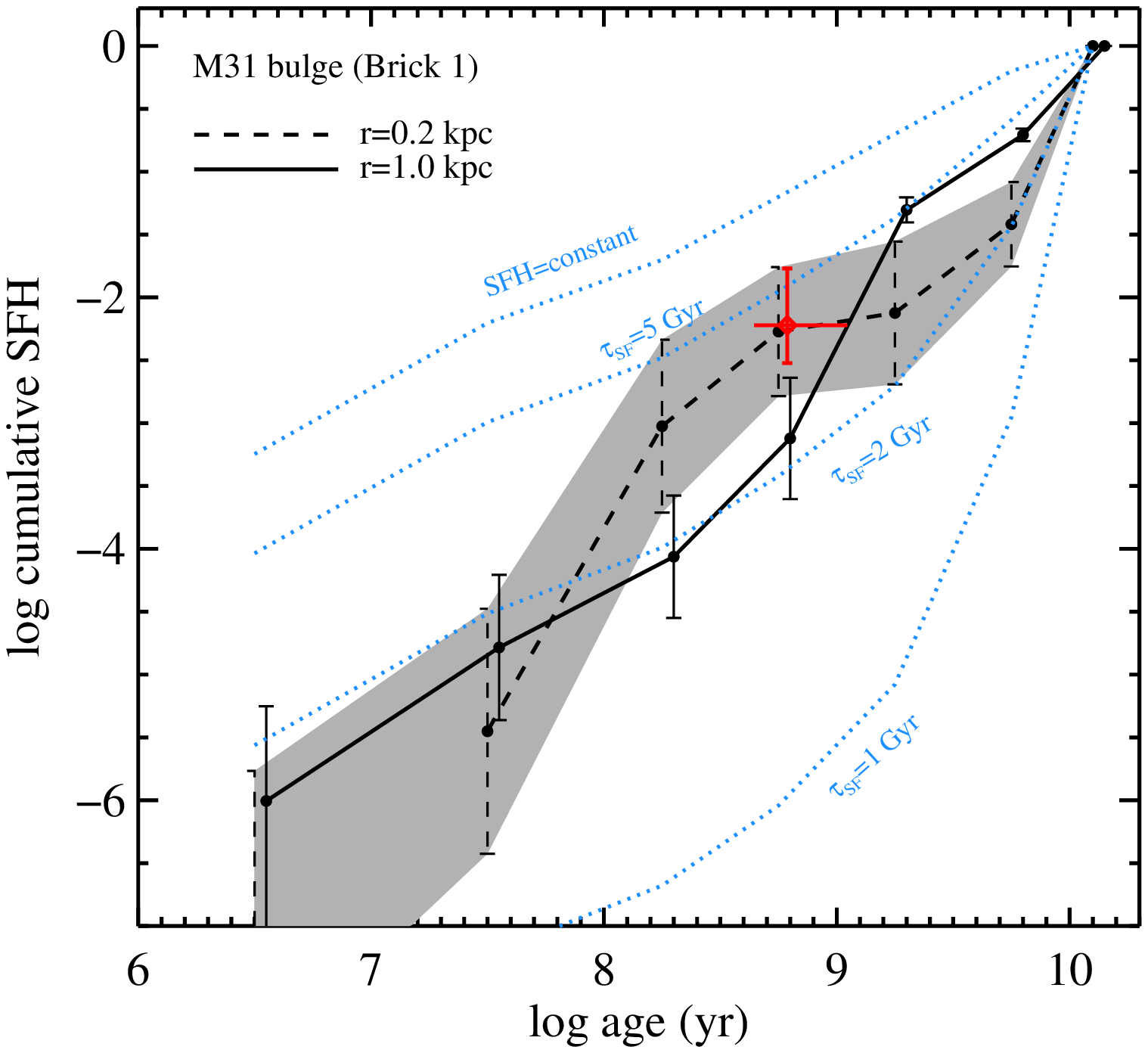}
\vspace{0.1cm}
\caption{{\it Left panels:} Comparison between pCMDs of two regions
  within the bulge of M31 (upper panels) and the corresponding
  best-fit models (lower panels).  The pCMDS are displayed as Hess
  diagrams with a logarithmic color mapping.  The best-fit $\npix$
  values are also shown in the legends of the bottom panels.  A 10 Gyr
  solar metallicity isochrone is shown in grey to guide the eye. {\it
    Right panel:} Cumulative SFH derived by fitting the observed pCMD
  to models.  The solid line is offset by 0.05 dex along the x-axis
  for clarity.  Error bars are $1\sigma$ uncertainties.  Also shown
  are several smooth SFHs to guide the eye (dotted lines).  The red
  point with error bars represents the mass fraction and age of the
  young stellar component reported by \citet{Dong15} from fitting
  integrated photometry for a region 0.2 kpc from the center.}
\label{fig:bulge}
\end{figure*}

\subsection{Tests With Mock Data}

Figure \ref{fig:fitmock} shows the results of several tests of SFH
recovery with mock data.  The mock data were constructed assuming a
solar metallicity population with zero reddening.  In each panel the
resulting best-fit cumulative SFH is shown as a solid line with the
grey bands marking the $1\sigma$ uncertainties.  The input SFH is also
shown as a dashed line.  While we focus here on the SFH recovery, we
are simultaneously also fitting for metallicity and dust content.  The
metallicities are recovered to within 0.05 dex, the reddening is
constrained to be $<0.01$, and the overall normalization, $\npix$ is
recovered to within 0.1 dex.

In the top panels we fit mock pCMDs generated with a constant SFH for
two values of $\npix$.  The upper left panel shows a model with
$\npix=10^2$.  The recovered SFH agrees very well with the input
value, with statistical uncertainties of less than a factor of two for
all age bins except the youngest where the uncertainties are a factor
of three.  In the upper right panel we show a model with $\npix=10^4$
and $100\times$ fewer pixels.  In other words, the total mass of the
two populations in the top panels are the same.  One can think of the
top panels as being of the same underlying system with the right panel
observed at a distance $10\times$ greater than the left panel.  As a
consequence, the information content is lower in the right panel
compared to the left and the uncertainties are therefore larger.
Nonetheless, even at $\npix=10^4$ one can reliably recover the full
SFH to remarkably high precision.  

The bottom left panel shows the result for an old stellar population
with an exponential decay time of $\tau_{\rm SF}=1$ Gyr and
$\npix=10^2$.  Here again the best-fit SFH agrees very well with the
input model.

The final test is shown in the bottom right panel of Figure
\ref{fig:fitmock}.  Here we take a constant SFH and increase the mass
in the second youngest bin by a factor of 5.  In other words, the
system has an underlying constant SFH with a recent, large burst of
star formation.  Here again the recovery is excellent, indicating that
we can recovery fairly detailed structure in the SFH of
crowding-limited systems by analyzing their pCMDs.

Overall we find these tests very encouraging as they imply that we can
reliably infer SFHs in a variety of astrophysically interesting
regimes including old and young stellar populations and populations
with bursty SFHs.  It is also encouraging that the best-fit solutions
and $1\sigma$ uncertainties encompass the input model in essentially
all age bins for all of our tests.

%---------------------------------------------------------%

\vspace{1cm}

\section{Comparison to Observations}
\label{s:data}

Having laid out the basic idea behind pCMDs and demonstrated that one
can {\it quantitatively} recover SFHs by fitting models to data in
pCMD space, we now turn to a comparison with observations.  For this
purpose we utilize {\it HST} observations of M31 obtained through the
PHAT survey \citep{Dalcanton12}.  Specifically, we use the brick-level
drizzled mosaics available in the MAST archive. We adopt a distance
modulus to M31 of 24.47 \citep{McConnachie05}.  In order to simulate
the effect of photon noise we adopt an exposure time in the $\gband$
and $\iband$ bands of 3620s and 3235s, respectively
\citep{Dalcanton12}.

\subsection{The Bulge of M31}

We first consider the old stellar population in the bulge of M31.  We
selected two regions from Brick 1 spanning a factor of $\approx10$ in
$\npix$.  The first region lies $\approx200$ pc from the center of M31
and has $\npix\sim10^3$ \citep[note that this is well beyond the inner
$\sim10$ pc where a young cluster of blue stars resides;][]{Lauer12}.
The second region lies 1 kpc from the center and has $\npix\sim10^2$.
Regions were selected over a relatively narrow range in radius from
the center in order to identify groups of pixels that would have a
relatively narrow distribution of $\npix$.  The first and second
regions include 253,000 and 79,000 pixels, respectively.  No attempt
was made to remove artifacts, clearly resolved bright stars, or other
features.  Note that the crowding limit is so severe in the bulge of
M31 that it is only possible to reliably photometer the most luminous
giants based on optical-NIR {\it HST} data \citep[][]{Dalcanton12,
  Williams14}.

\begin{figure*}[!t]
\center
\plottwo{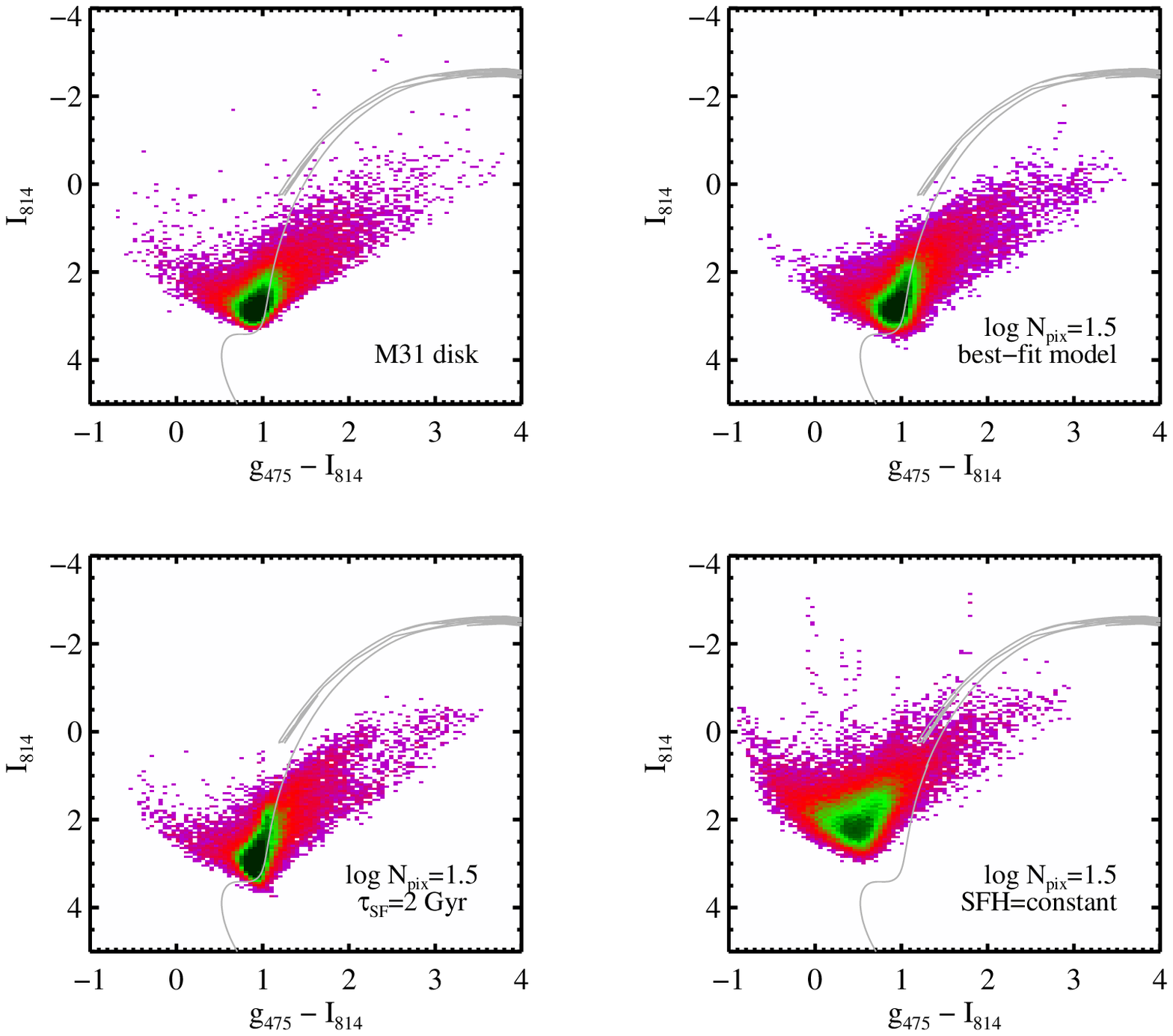}{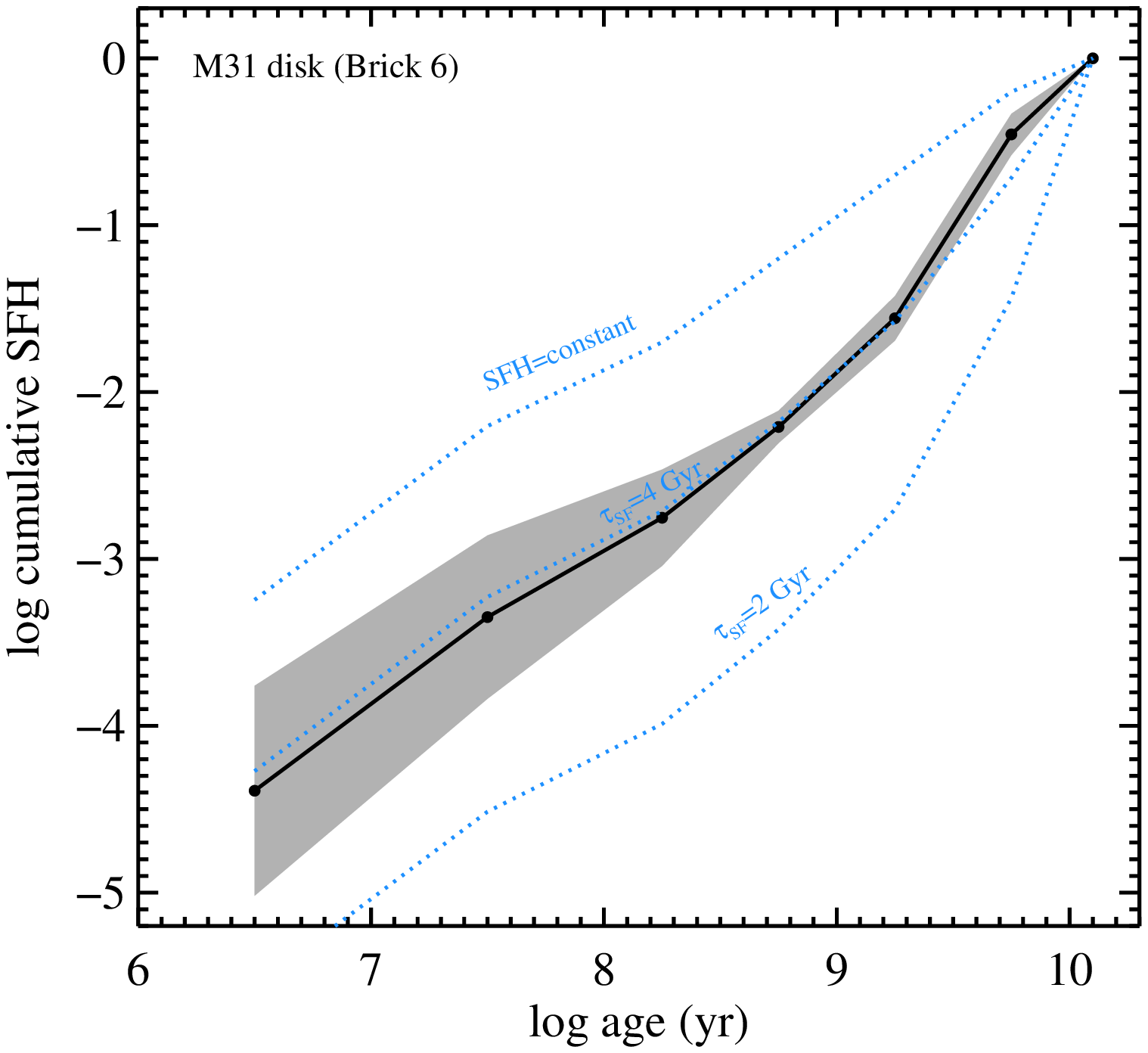}
\vspace{0.1cm}
\caption{{\it Left panels:} Comparison between pCMDs of a small region
  of the star-forming disk of M31 (upper left) and models with
  $\npix=10^{1.5}$.  This particular region was chosen to coincide
  with one of the regions used in resolved star CMD fitting (see
  Figure \ref{fig:fitcompare}). The models include the best-fit (upper
  right) and simple, solar metallicity models with a constant SFH and
  a $\tau$-model SFH (bottom panels).  The pCMDs are displayed as Hess
  diagrams with a logarithmic color mapping.  A 10 Gyr solar
  metallicity isochrone is shown in grey to guide the eye. {\it Right
    panel:} Cumulative SFH derived by fitting the observed pCMD to
  models.   Shaded region represents the $1\sigma$ uncertainties.
  Also shown are several smooth SFHs to guide the eye (dotted lines).}
\label{fig:disk}
\end{figure*}

The left panels of Figure \ref{fig:bulge} show a comparison of pCMDs
between these two regions of the bulge of M31 and the best-fit models.
The effect of the PSF and observational (photon counting)
uncertainties are included in the models.  Overall the best-fit models
do a good job of reproducing the features in the observed pCMDs.  In
detail the observations appear to span a slightly wider range in
colors at a fixed luminosity.  This may be pointing to the fact that
our models are too simplistic.  For example, allowing for a
metallicity distribution function would result in a broader range of
colors at fixed luminosity.  We defer these complications to future
work.

The right panel of Figure \ref{fig:bulge} shows the derived cumulative
SFHs for these two regions after fitting the observed pCMDs to models.
Also shown are simple SFHs to guide the eye.  The derived SFHs are
steeply declining with time and are broadly consistent with a
$\tau-$model SFH with $\tau_{\rm SF}\approx2$ Gyr.  It is also
interesting to note that the recovered SFHs are not declining as fast
as possible, i.e., a $\tau_{\rm SF}=1$ Gyr model appears to be ruled
out by the data.  This is noteworthy in light of the integrated light
analysis presented in \citet{Dong15}.  These authors modelled FUV-NIR
photometry in the bulge of M31 and presented evidence for a young
stellar population with an age of $\sim600-800$ Myr and a mass
fraction of $\sim1-2$\% in the inner 700 pc.  We show their derived
age and mass fraction for the young component in the right panel of
Figure \ref{fig:bulge} for their $50-55$\arcsec bin, which corresponds
closely to our 0.2 kpc region.  Their result agrees very well with our
derived SFH for this region.

The existence of hot stars associated with old stellar populations
(including blue stragglers hot horizontal branch stars, and post-AGB
stars) has long complicated the measurement of low levels of SF in
such systems, as these hot evolved stars can, in certain
circumstances, masquerade as young main sequence stars.  Such stars,
if present in significant numbers in the bulge of M31, could also bias
our derived SFHs high at young and intermediate ages.  The most
conservative interpretation of our results is that they represent
upper limits, as some of the bluest pixels could be due to the hot
evolved stars not currently included in our models.  We will explore
the effect of hot evolved stars on the derived SFHs from pCMDs if
future work.

The best-fit metallicities are close to solar and the reddening values
are $E(B-V)\lesssim0.01$.  There is some age-dust degeneracy.  In
light of Figure \ref{fig:mpix_z} it would be preferable to use an
optical-NIR color in order to more strongly separate these two
variables.  Nonetheless, the modest degeneracy between age and
metallicity does not have a significant impact on the recovery of the
SFH.

\subsection{The Star-Forming Disk of M31}

We now consider the stellar population in the star forming disk of
M31.  For this comparison we selected a single $\approx100$ pc
$\times$ 100 pc region from Brick 6 data.  The region corresponds to
one of the 9000 regions analyzed by \citet{Lewis15}, who used the
resolved stars in this region to constrain the SFH over the past $500$
Myr.  The resulting pCMD is shown in the upper left panel of Figure
\ref{fig:disk}.  We compare these observed pCMD to three models: the
best-fit model, and two simple models.  The latter two models are
solar metallicity, reddening-free, and have a constant SFH and a
$\tau-$model SFH with $\tau_{\rm SF}=2$ Gyr.  These simple models were
generated with $\npix=10^{1.5}$, which is very close to the best-fit
$\npix$ value.

The right panel of Figure \ref{fig:disk} shows the derived cumulative
SFH for this region.  We also show several simple SFHs in order to
guide the eye.  The derived SFH agrees remarkably well with a smooth
exponential model with $\tau_{\rm SF}=4$ Gyr.  The best-fit
metallicity is close to solar and the reddening is low,
$E(B-V)\sim0.01$.

\begin{figure*}[!t]
\center
\includegraphics[width=0.94\textwidth]{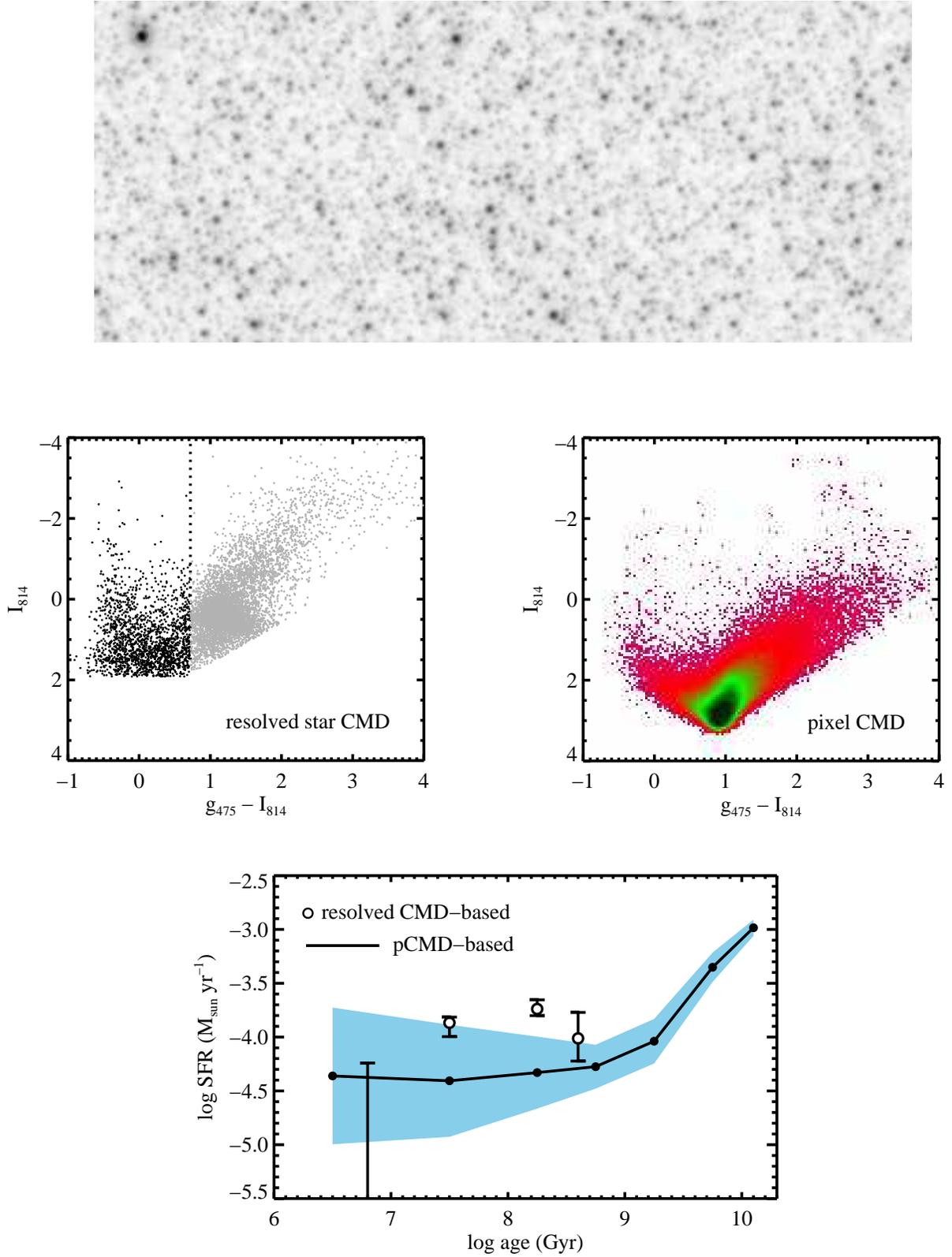}
\vspace{0.1cm}
\caption{Direct comparison between resolved and pixel CMDs, and SFHs
  derived therefrom.  {\it Top panel:} $\iband-$band image of a
  $40\times90$ pc subregion of the full $100\times100$ pc region used
  in the analysis.  {\it Middle left panel:} CMD of resolved sources.
  The resolved sources are truncated at $\iband\approx2$ which
  corresponds to the 50\% completeness limit of the catalog (fainter
  stars are crowding-limited).  Stars redder than the dotted line were
  excluded from the resolved star CMD analysis in \citet{Lewis15}.
  {\it Middle right panel:} Pixel CMD, represented as a Hess diagram
  with a logarithmic color map.  {\it Bottom panel:} SFHs derived from
  the resolved star CMD (points) and from the pCMD (solid line).  The
  shaded region and error bars represent $1\sigma$ uncertainties.  The
  resolved star SFH was modeled with the exact same age binning used
  for the pCMD-based SFHs.  The overall agreement is encouraging given
  the different systematic uncertainties between the two techniques.
  The resolved CMD-based SFH was derived from fitting the main
  sequence and so is limited to younger ages because the main sequence
  turnoff point for older ages is below the crowding limit. }
\label{fig:fitcompare}
\end{figure*}

The region that we have analyzed was chosen so that a direct
comparison could be made with the resolved star analysis in
\citet{Lewis15}.  The results are shown in Figure
\ref{fig:fitcompare}.  The top panel shows an $\iband-$band image of a
$40\times90$ pc subregion of the full $100\times100$ pc region used in
the analysis\footnote{These are projected distances; the de-projected
  regions are approximately $100\times450$ pc.}. The middle panels
show the resolved star CMD and the pixel CMD for the full
$100\times100$ pc region.  \citet{Lewis15} fit the region of the
resolved CMD blueward of the dotted line in order to focus on the main
sequence, which is easier to model than the evolved giants.  There are
1800 stars in the fitted region of the diagram.  The sharp cutoff at
$\iband\approx2$ corresponds to the 50\% completeness limit of the
resolved star catalog \citep{Lewis15}.

The derived SFHs from these two approaches are compared in the bottom
panel of Figure \ref{fig:fitcompare}.  The resolved star SFH from
Lewis et al. is based on modeling the main sequence and hence is
limited to the most recent $\approx500$ Myr; the main sequence turnoff
is below the crowding limit for older stellar populations
\citep[see][for SFHs derived to older ages in M31 based on modeling
the evolved giants]{Williams15}.  The $1\sigma$ uncertainties on the
resolved star SFHs are statistical only; at the youngest age they are
dominated by Poisson uncertainties in the number of stars above the
crowding limit.  The resolved star SFH was re-computed specifically
for this comparison for the exact same age bins as the pixel CMDs
(courtesy of A. Lewis), and the results were multiplied by 1.5 to
convert from a \citet{Kroupa01} IMF to our adopted Salpeter IMF.

The overall agreement is encouraging.  Three of the four age bins
agree very well within the errors.  The third age bin differs more
significantly.  In addition to the overall differences in techniques,
there are a variety of issues that could drive these differences.
First and foremost, we adopt a different set of stellar evolution
models than those in \citet{Lewis15}, who adopt the Padova models.  It
would be interesting to repeat both analysis with a common set of
stellar evolution models, as different isochrone tables can induce
non-trivial systematics \citep[e.g.,][]{Dolphin12, Weisz14}.  Second,
\citet{Lewis15} adopt a more sophisticated dust model than employed
herein.  They consider a model that allows for a distribution of dust
that includes foreground extinction and differential extinction; the
extinction PDF is a step function between the foreground and
differential extinction values.  For this particular region they find
a foreground extinction of $E(B-V)=0.1$ (assuming $R_V=3.1$) and a
differential extinction of $E(B-V)=0.23$.  Our best-fit reddening is
very low by comparison: $E(B-V)\sim0.01$.  We do not yet know what is
responsible for driving our reddening values so low, but it could be
related to our simplified treatment of reddening (one value applied to
all stars equally).  In future work we will explore the more
sophisticated dust model employed in \citet{Lewis15}.  However, we
have tested the impact of our low derived reddening values on the
inferred SFH.  We re-ran the fitting with a lower limit on the
$E(B-V)$ prior of 0.08.  The resulting SFRs agree with those shown for
our default model in \ref{fig:fitcompare} within $1\sigma$ even though
the best-fit $E(B-V)$ is closer to 0.1.  This test suggests that the
derived SFHs are not overly sensitive to the best-fit reddening.
Another potential systematic uncertainty is our use of a PSF that does
not take into account the effects of the drizzling process that was
used to create the final mosaic images.

As a further test of the pCMD-based results, we have used our best-fit
parameters to predict the integrated fluxes within this region from
the FUV through NIR.  For the F475W, F814W, F110W, and F160W filters
our predicted fluxes agree to within 3\%, 8\%, 13\%, and 15\%,
respectively.  We have also synthesized FUV and NUV fluxes from GALEX
images and find that our predicted fluxes are about a factor of 3 too
bright.  If we adopt a reddening of $E(B-V)=0.13$ then the model and
observed fluxes are in much better agreement, providing additional
support to the conclusion that our best-fit reddening values are too
low.

Finally, we note that the metallicities derived in the two techniques
agree well and both return metallicities within $\approx0.05$ dex of
solar.

We summarize this comparison by concluding that the broad agreement
between the resolved star and pCMD-based fitting for the SFH and
metallicity is encouraging, but that our model for dust is likely too
simplistic.  Preliminary tests suggest that this shortcoming does not
unduly bias our derived SFH but further tests are needed.

%---------------------------------------------------------%

\section{Discussion}
\label{s:disc}

We envision pCMDs being employed in at least two related but distinct
regimes.  First, in the regime where resolved photometry is not
possible even for the brightest stars, pCMDs offers a unique
opportunity to extract stellar population information.  We presented
one example here for the bulge of M31.  Other examples include more
distant systems (e.g., massive galaxies in the Virgo cluster).  As we
have remarked earlier, the pCMD concept is in some sense a
generalization of the SBF technique, which has been used both as a
distance indicator and as a probe of stellar populations
\citep[e.g.,][]{Tonry01, Cantiello05, Blakeslee10}.  By analogy with
the SBF technique, one could imagine including distance as an
additional free parameter when fitting pCMDs.  Distance results in an
overall vertical shift in the pCMD, and no other parameter induces a
similar vertical shift in the pCMD.  For this reason we anticipate
that pCMDs will deliver strong constraints on the distance.

Second, in the regime in which bright stars are resolvable but the
oldest main sequence turnoff point remains below the crowding limit,
as is the case in the optical-NIR throughout the disk of M31, pCMDs
offer a valuable complementary tool that can be used in combination
with classic resolved star techniques.  In this regime the key point
is that for a given age, if the crowding limit is above the turnoff
point then there is additional age information in between the crowding
limit and the turnoff point that is not used in traditional resolved
star analysis but is contained in pCMDs.  For example,
\citet{Williams07} used the surface brightness below the magnitude
limit of their resolved star data to place additional constraints on
the SFH of intracluster stars in the Virgo cluster.  As an example of
the possible synergy between resolved star and pCMD analyses, the dust
model from the resolved star analysis as a prior on the dust model in
the pCMD analysis.  Ultimately, one could imagine jointly fitting the
resolved CMD and pCMD data within the same model, thereby providing
the strongest possible constraints on physical parameters of the
system.

We emphasize that pCMDs provide a much less obstructed view of the
main sequence than is available from standard integrated light
observations.  For example, for an old stellar population viewed in
integrated light, the main sequence comprises $50-60$\% of the light
at $0.5-0.6\mu m$, and $\approx30$\% at $0.8\mu m$
\citep[e.g.,][]{Conroy13b}.  We computed similar numbers for our pCMDs
as a function of pixel luminosity, for the case of a $\tau_{\rm SF}=1$
Gyr SFH and $\npix=10^2$.  At the faintest pixels ($\iband\approx2$),
the main sequence contributes 80\% of the flux at $0.8\mu m$.  The key
point is that the pCMDs are sensitive both to the evolved giants and
to the sea of main sequence stars, and with pCMDs we can separately
extract the age and metallicity-sensitive information from these
components.  Moreover, by combining pCMDs with spectroscopy at the
pixel level one can hope to extract even more information
\citep[e.g.,][]{Mould12, vanDokkum14}.

The majority of ground and space-based facilities coming online in the
next two decades will not deliver substantially increased angular
resolution compared to current facilities, and as a consequence they
will not improve upon the crowding limit offered by {\it HST}.  The
ELTs do offer the hopes of dramatically increased angular resolution.
In this case many more galaxies will be in the semi-resolved regime
(e.g., the central regions of galaxies in the Virgo cluster).  In
light of this, we believe that one must embrace crowding-limited data
and develop tools for extracting information in that regime.  pCMDs
offer one such example in this direction.

This is the first attempt to fit observations in pCMD space, and, as a
consequence, a number of simplifying assumptions have been made.  Due
to computational limitations we have made several shortcuts; we are
optimistic that solutions to these limitations can be found in the
near future.  Our underlying model is relatively simple, containing
only 9 parameters (7 for the SFH, and one each for metallicity and
reddening).  Clearly the reddening model is too simplistic and it
should be straightforward to expand this component into something that
rivals the resolved star analysis in sophistication \citep[see
e.g.,][for the current state of the art]{Dalcanton15}.  Likewise for
the metallicity, in principle it is straightforward to add more
components, but in this case the computational burden increases
linearly with each added component.  Greater care needs to be taken in
handling the (spatially variable) PSF.  The relative simplicity of the
model also suggests that our quoted uncertainties are likely lower
limits.  Nonetheless, there are no obvious ``show stoppers'' in the
modeling of pCMDs, and so we encourage their use for interpreting
crowding-limited data.

%---------------------------------------------------------%

\section{Summary}
\label{s:sum}

We have presented the concept of pixel color magnitude diagrams
(pCMDs) as a powerful tool for analyzing stellar populations in the
crowding-limited regime.  We constructed stellar population models and
highlighted the main dependencies on the key parameters.  We also
compared the model pCMDs to {\it HST} imaging of M31 obtained through
the PHAT survey.  We now summarize our main results.

\begin{itemize}

\item A key parameter governing the behavior of pCDMs is the mean
  number of stars per pixel, $\npix$.  This parameter governs how
  mottled or smooth the image appears, and when combined with the
  measured flux, can provide a strong constraint on the distance to
  the system.  This is not surprising as the ``magnitude'' part of
  pCMDs is essentially equivalent to the information contained in
  SBFs.

\item pCMDs show strong sensitivity to the underlying SFH with the
  ability to resolve bursts of star formation and place strong
  constraints on old stellar populations even when the data are
  strongly crowding-limited.  Our simulations have demonstrated that
  one can recover at least 7 age components non-parametrically with
  only two filter data.
  
\item In addition to the SFH, the metallicity and dust content can
  also be reliably separated with pCMDs, especially with optical-NIR
  colors.  While both metallicity and dust result in redder colors,
  the detailed structure of the data in pCMD space offers strong,
  separable constraints on both the metallicity and the overall
  reddening.  Again, all of this can be measured with only two filter
  data.

\item We have developed machinery to fit model pCMDs to observations.
  We then constructed pCMDs from {\it HST} imaging of M31 in several
  small regions in the bulge and disk.  We derived non-parametric SFHs
  in these regions extending from $10^6$ yr to $10^{10}$ yr.  These
  are the strongest constraints to-date on the full SFH in the bulge
  of M31.

\item We also compared our SFHs to those derived from fitting the
  resolved star CMD in one disk field and found overall good agreement
  for ages $10^{6.5}-10^{8.7}$ yr, where the resolved star results
  were available.  The comparison revealed notable disagreement in
  the derived reddening, suggesting that there are important
  systematic uncertainties in our current approach to fitting pCMDs.
  Going forward, the combination of resolved star and pCMD data should
  provide the strongest possible constraints on the full SFH.

\end{itemize}

\acknowledgments 

We thank Julianne Dalcanton, Ben Johnson, Alexia Lewis, Phil
Rosenfield and Dan Weisz for helpful discussions throughout the
development of this work and for comments on an earlier draft, Jieun
Choi for providing custom-made isochrones, Alexia Lewis for providing
the custom-made PHAT SFHs and resolved star photometry, and Dan
Foreman-Mackey for advise on statistical matters. C.C. acknowledges
support from NASA grant NNX13AI46G, NSF grant AST-1313280, and the
Packard Foundation.  The computations in this paper were run on the
Odyssey cluster supported by the FAS Division of Science, Research
Computing Group at Harvard University.

%\bibliography{../master_refs}

\end{document}